\documentstyle[12pt]{article}

\begin{document}
\begin{titlepage}
\begin{center}

May 14, 1999     \hfill    LBNL-42650 \\

\vskip .3in

{\large \bf Attention, Intention, and Will in Quantum Physics}
\footnote{This work was supported by the Director, Office of Energy 
Research, Office of High Energy and Nuclear Physics, Division of High 
Energy Physics of the U.S. Department of Energy under Contract 
DE-AC03-76SF00098.}
\vskip .30in
Henry P. Stapp\\
{\em Lawrence Berkeley National Laboratory\\
      University of California\\
    Berkeley, California 94720}
\end{center}

\vskip .2in

\begin{abstract}

How is mind related to matter? This ancient question in philosophy is rapidly
becoming a core problem in science, perhaps the most important of all because 
it probes the essential nature of man himself. The origin of the
problem is a conflict between the mechanical conception of human beings 
that arises from the precepts of classical physical theory and the very 
different idea that arises from our intuition: the former reduces each of us 
to an automaton, while the latter allows our thoughts to guide our actions.
The dominant contemporary approaches to the problem attempt to resolve this 
conflict by clinging to the classical concepts, and trying to explain away 
our misleading intuition. But a detailed argument given here shows why, in a 
scientific approach to this problem, it is necessary to use the more basic 
principles of quantum physics, which bring the observer into the dynamics, 
rather than  to accept classical precepts that are profoundly incorrect 
precisely at the crucial point of the role of human consciousness in 
the dynamics of human brains. Adherence to the quantum principles
yields a dynamical theory of the mind/brain/body system that is in close
accord with our intuitive idea of what we are. In particular, the need for
a self-observing quantum system to pose certain questions creates a causal
opening that allows mind/brain dynamics to have three distinguishable but
interlocked causal processes,  one micro-local, one stochastic, and the third
experiential.  The classical approximation reduces this tripartite quantum 
process to a single deterministic local process: setting Planck's constant to
zero eliminates the dynamical fine structure wherein the effect of mind on 
matter lies.

\end{abstract} 
\medskip
\end{titlepage}

\renewcommand{\thepage}{\roman{page}}
\setcounter{page}{2}
\mbox{ }

\vskip 1in

\begin{center}
{\bf Disclaimer}
\end{center}

\vskip .2in

\begin{scriptsize}
\begin{quotation}
This document was prepared as an account of work sponsored by the United
States Government. While this document is believed to contain correct 
 information, neither the United States Government nor any agency
thereof, nor The Regents of the University of California, nor any of their
employees, makes any warranty, express or implied, or assumes any legal
liability or responsibility for the accuracy, completeness, or usefulness
of any information, apparatus, product, or process disclosed, or represents
that its use would not infringe privately owned rights.  Reference herein
to any specific commercial products process, or service by its trade name,
trademark, manufacturer, or otherwise, does not necessarily constitute or
imply its endorsement, recommendation, or favoring by the United States
Government or any agency thereof, or The Regents of the University of
California.  The views and opinions of authors expressed herein do not
necessarily state or reflect those of the United States Government or any
agency thereof or The Regents of the University of California and shall
not be used for advertising or product endorsement purposes.
\end{quotation}
\end{scriptsize}

\vskip 2in

\begin{center}
\begin{small}
{\it Lawrence Berkeley Laboratory is an equal opportunity employer.}
\end{small}
\end{center}

\newpage
\renewcommand{\thepage}{\arabic{page}}
\setcounter{page}{1}

\begin{center}
{\bf Shifting the Paradigm}
\end{center}
\vspace{.1in}

A controversy is raging today about the power of our minds. Intuitively we 
know that our conscious thoughts can guide our actions. Yet 
the chief philosophies of our time proclaim, in the name of science, 
that we are mechanical systems governed,  fundamentally, entirely by 
impersonal laws that operate at the level of our  microscopic constituents.  

The question of the nature of the relationship between conscious thoughts 
and physical actions is called the mind-body problem.  Old as 
philosophy itself it was brought to its present form by the 
rise, during the seventeenth century, of what is 
called `modern science'. The ideas of Galileo Galilei, 
Ren\^{e} Descartes, and Isaac Newton created a magnificent 
edifice known as classical physical theory, which was 
completed by the work of James Clerk Maxwell and Albert
Einstein. The central idea is that the physical universe is composed 
of ``material'' parts that are localizable in tiny regions, and that 
all motion of matter is completely determined by matter alone, via 
local universal laws. This {\it local} character 
of the laws is crucial. It means that each tiny 
localized part responds only to the states of its immediate 
neighbors: each local part ``feels'' or ``knows about'' 
nothing outside its immediate microscopic neighborhood. 
Thus the evolution of the physical universe, and of every 
system within the physical universe, is governed by a vast 
collection of local processes, each of which is `myopic' 
in the sense that it `sees' only its immediate neighbors.

The problem is that if this  causal structure 
indeed holds then there is no need for our human feelings and 
knowings. These experiential qualities clearly correspond to 
large-scale properties of our brains. But if the entire 
causal process is already completely determined by 
the `myopic' process postulated by classical physical theory,
then there is nothing for any unified graspings of 
large-scale properties to do. Indeed, there is nothing that 
they {\it can} do that is not already done by the myopic 
processes. Our conscious thoughts thus become prisoners of 
impersonal microscopic processes: we are, according to this 
``scientific'' view, mechanical robots, with a mysterious
dangling appendage, a stream of conscious thoughts that can grasp
large-scale properties as wholes, but exert, as a consequence of these
graspings, nothing not done already by the microscopic constituents.

The enormous empirical success of classical physical theory 
during the eighteenth and nineteenth centuries has led many 
twentieth-century philosophers to believe that the problem with 
consciousness is how to explain it away: how to discredit our misleading 
intuition by identifying it as product of human confusion, rather than 
recognizing the physical effects of consciousness as a physical problem 
that needs to be answered in dynamical terms. That strategy of evasion is, 
to be sure, about the only course available within the strictures imposed 
by classical physical theory. 

Detailed proposals abound for how to deal with this problem 
created by adoption of the classical-physics world view. 
The influential philosopher Daniel Dennett (1994, p.237) 
claims that our normal intuition about consciousness is 
``like a benign user illusion'' or ``a metaphorical 
by-product of the way our brains do their approximating 
work''. Eliminative materialists such as Richard Rorty (1979) 
hold that mental phenomena, such as conscious experiences, 
simply do not exist. Proponents of the popular `Identity Theory of 
Mind' grant that conscious experiences do exist, but claim 
each experience to be {\it identical} to some brain process. 
Epiphenomenal dualists hold that our conscious experiences 
do exist, and are not identical to material processes, but 
have no effect on anything we do: they are epiphenomenal.

Dennett (1994, p.237)  described the recurring idea that pushed him to his
counter-intuitive conclusion: ``a brain was always going to do what it was 
caused to do by local mechanical disturbances.'' This passage lays bare  
the underlying presumption behind his own theorizing, and undoubtedly behind
the theorizing of most non-physicists who ponder this matter, namely 
the presumptive essential correctness of the idea of the physical world 
foisted upon us by the assumptions of classical physical theory.

It has become now widely appreciated that assimilation by the general public
of this ``scientific'' view, according to which each human being is basically 
a mechanical robot, is likely to have a significant and corrosive impact on 
the moral fabric of society. Dennett speaks of the Spectre of Creeping 
Exculpation: recognition of the growing tendency of people to exonerate 
themselves by arguing that it is not ``I'' who is at fault, but some 
mechanical process within: ``my genes made me do it'';  or ``my high 
blood-sugar content made me do it.'' [Recall the infamous ``Twinkie Defense'' 
that got Dan White off with five years for murdering San Francisco Mayor 
George Moscone and Supervisor Harvey Milk.] 

Steven Pinker (1997, p.55) also defends a classical-type conception
of the brain, and, like Dennett, recognizes the important need to 
reconcile the science-based idea of causation with a rational conception 
of personal responsibility. His solution is to regard science and ethics 
as two self-contained systems: ``Science and morality are separate spheres
of reasoning. Only by recognizing them as separate can we have them both.'' 
And ``The cloistering of scientific and moral reasoning also lies behind my 
recurring metaphor of the mind as machine, of people as robots.'' But he then 
decries ``the doctrines of postmodernism, poststructuralism, and 
deconstructionism, according to which objectivity is impossible, meaning is 
self-contradictory, and reality is socially constructed.'' Yet are not the 
ideas he decries a product of the contradiction he embraces? 
Self-contradiction is a bad seed that bears relativism as its evil fruit.  

The current welter of conflicting opinion about the mind-brain connection
suggests that a paradigm shift is looming. But it will require a major 
foundational shift. For powerful thinkers have, for three centuries, been 
attacking this problem  from every angle within the bounds defined by the 
precepts of classical physical theory, and no consensus has emerged.

Two related developments of great potential importance are now 
occurring. On the experimental side, there is an explosive 
proliferation of empirical studies of the relations between a subject's 
brain process --- as revealed by instrumental probes of diverse kinds --- 
and the experiences he reports. On the theoretical side, there is a 
growing group of physicists who believe almost all thinking on this 
issue during the past few centuries to be logically unsound, because it is 
based implicitly on the precepts of classical physical theory, which are now  
known to be fundamentally incorrect. Contemporary physical theory differs
profoundly from classical physical theory precisely on the nature of 
the dynamical linkage between minds and physical states.
	
William James (1893, p.486),  writing  at the end of the nineteenth century,
said of the scientists who would one day illuminate the mind-body problem:

``the best way in which we can facilitate their advent is to understand how 
great is the darkness in which we grope, and never forget that the 
natural-science assumptions with which we started are provisional and 
revisable things.''

How wonderfully prescient!

It is now well known that the precepts of classical physical theory are 
fundamentally incorrect. Classical physical theory has been superceded by 
quantum theory, which reproduces all of the empirical successes of classical 
physical theory, and succeeds also in every  known case where the predictions 
of classical physical theory fail. Yet even though quantum theory yields all 
the correct predictions of classical physical theory, its representation of 
the physical aspects of nature is profoundly different from that of 
classical physical theory. And the most essential difference 
concerns precisely the connection between physical states and consciousness. 

My thesis here is that the difficulty with the traditional attempts to
understand the mind-brain system lies primarily with the physics 
assumptions, and only secondarily with the philosophy:  once the physics 
assumptions are rectified the philosophy will take care of itself. A correct 
understanding of the mind/matter connection cannot be based on a 
conception of the physical aspects of nature that is profoundly mistaken 
precisely at the critical point, namely the role of consciousness in the 
dynamics of physical systems.

Contemporary science, rationally pursued, provides an essentially new 
understanding of the mind/brain system. This revised understanding is in 
close accord with our intuitive  understanding of that system: no idea of 
a ``benign user illusion'' arises,  nor any counter-intuitive idea that a 
conscious thought is identical to a collection of tiny objects moving about 
in some special kind of way. 

Let it be said, immediately, that this solution lies not in the invocation 
of quantum randomness: a significant dependence of human action on random 
chance would be far more destructive of any rational notion of personal 
responsibility than microlocal causation ever was. 

The solution hinges not on quantum randomness, but rather on the dynamical
effects within quantum theory of the intention and attention of the observer.

But how did physicists ever manage to bring conscious thoughts into 
the dynamics of physical systems? That is an interesting tale.

\vspace{.2in}
\begin{center}
                    {\bf  The World as Knowings}
\end{center} 
\vspace{.1in}

In his book ``The creation of quantum mechanics and the Bohr- Pauli
dialogue" the historian John Hendry (1984) gives a detailed
account of the fierce struggles, during the first quarter of this
century, by such eminent thinkers as Hilbert, Jordan, Weyl, von Neumann,
Born, Einstein, Sommerfeld, Pauli, Heisenberg, Schroedinger, Dirac,
Bohr and others, to come up with a rational way of comprehending the
data from atomic experiments.  Each man had his own bias and
intuitions, but in spite of intense effort no rational comprehension
was forthcoming.  Finally, at the 1927 Solvay conference a group
including Bohr, Heisenberg, Pauli, Dirac, and Born come into
concordance on a solution that came to be called ``The Copenhagen
Interpretation".  Hendry says: ``Dirac, in discussion, insisted on the
restriction of the theory's application to our knowledge of a system,
and on its lack of ontological content."  Hendry summarized the
concordance by saying: ``On this interpretation it was agreed that, as
Dirac explained, the wave function represented our knowledge of the
system, and the reduced wave packets our more precise knowledge after
measurement."

Let there be no doubt about this key point, namely that the
mathematical theory was asserted to be directly about our knowledge
itself, not about some imagined-to-exist world of particles and
fields.

Heisenberg (1958a): ``The conception of objective reality of the
elementary particles has thus evaporated not into the cloud of some
obscure new reality concept but into the transparent clarity of a
mathematics that represents no longer the behavior of particles but
rather our knowledge of this behavior."

Heisenberg (1958b): ``...the act of registration of the result in the
mind of the observer.  The discontinuous change in the probability
function...takes place with the act of registration, because it is the
discontinuous change in our knowledge in the instant of registration
that has its image in the discontinuous change of the probability
function."

Heisenberg (1958b:) ``When the old adage `Natura non facit saltus' is 
used as a basis of a criticism of quantum theory, we can reply that
certainly our knowledge can change suddenly, and that this fact
justifies the use of the term `quantum jump'. "  

Wigner (1961): ``the laws of quantum mechanics cannot be
formulated ... without recourse to the concept of consciousness."

Bohr (1934): ``In our description of nature the purpose is not to
disclose the real essence of phenomena but only to track down as far
as possible relations between the multifold aspects of our
experience."

Certainly this profound shift in physicists' conception of the basic
nature of their endeavor, and the meanings of their formulas, was not
a frivolous move: it was a last resort.  The very idea that in order
to comprehend atomic phenomena one must abandon ontology, and construe
the mathematical formulas to be directly about the knowledge of human
observers, rather than about the external real events themselves, is
so seemingly preposterous that no group of eminent and renowned
scientists would ever embrace it except as an extreme last measure.
Consequently, it would be frivolous of us simply to ignore a
conclusion so hard won and profound, and of such apparent direct
bearing on our effort to understand the connection of our knowings to
our physical actions.

This monumental shift in the thinking of scientists was an epic event
in the history of human thought.  Since the time of the ancient Greeks
the central problem in understanding the nature of reality, and our
role in it, has been the puzzling separation of nature into two
seemingly very different parts, mind and matter.  This had led to the
divergent approaches of Idealism and Materialism.  According to the
precepts of Idealism our ideas, thoughts, sensations, feelings, and
other experiential realities, are the only realities whose existence
is certain, and they should be taken as basic.  But then the enduring
external structure normally imagined to be carried by matter is
difficult to fathom.  Materialism, on the other hand, claims that
matter is basic.  But if one starts with matter then it is difficult
to understand how something like your experience of the redness of a
red apple can be constructed out of it, or why the experiential aspect
of reality should exist at all if, as classical mechanics avers, the
material aspect is causally complete by itself.  There seems to be no
rationally coherent way to comprehend the relationship between our
thoughts and the thoughtless atoms that external reality was imagined
to consist of.

Einstein never accepted the Copenhagen interpretation. He said: 

``What does not satisfy me, from the standpoint of principle, is its
attitude toward what seems to me to be the programmatic aim of all
physics: the complete description of any (individual) real situation
(as it supposedly exists irrespective of any act of observation or
substantiation)." (Einstein, 1951, p.667)

and 

``What I dislike in this kind of argumentation is the basic
positivistic attitude, which from my view is untenable, and which
seems to me to come to the same thing as Berkeley's principle, esse
est percipi."  (Einstein, 1951, p.  669).[Translation: To be is to be
perceived]

Einstein struggled until the end of his life to get the observer's
knowledge back out of physics.  But he did not succeed!  Rather he
admitted that: 

``It is my opinion that the contemporary quantum
theory...constitutes an optimum formulation of the [statistical]
connections."  (ibid.  p. 87).  

He referred to: 

``the most successful physical theory of our period, viz., the
statistical quantum theory which, about twenty-five years ago took on
a logically consistent form.  ... This is the only theory at present
which permits a unitary grasp of experiences concerning the quantum
character of micro-mechanical events." (ibid p. 81).  

One can adopt the cavalier attitude that these profound difficulties
with the classical conception of nature are just some temporary
retrograde aberration in the forward march of science. Or one can
imagine that there is simply some strange confusion that has
confounded our best minds for seven decades, and that their absurd
findings should be ignored because they do not fit our intuitions. Or one
can try to say that these problems concern only atoms and molecules,
and not things built out of them.  In this connection Einstein said:

``But the `macroscopic' and `microscopic' are so inter-related that it
appears impracticable to give up this program [of basing physics on
the `real'] in the `microscopic' alone." (ibid, p.674).

\vspace{.2in}
\begin{center}
{\bf                    What Is Really Happening?     } 
\end{center}
\vspace{.1in}

Orthodox quantum theory is pragmatic: it is a practical tool based on
human knowings.  It takes our experiences as basic, and judges theories 
on the basis of how well they work {\it for us}, without trying to attribute 
any reality to the entities of the theory, beyond the reality {\it for us} 
that they acquire from their success in allowing us to find rational order 
in the structure of our past experiences, and to form sound expectations about 
the consequences of our possible future actions.

But the opinion of many physicists, including Einstein, is that the proper 
task of scientists is to try to construct a rational theory of nature that is 
not based on so small a part of the natural world as human knowledge.
John Bell opined that we physicists ought to try to do better than that.

The question thus arises as to what is `really happening'.

Heisenberg (1958) answered this question in the following way: 

``Since through the observation our knowledge of the system has 
changed discontinuously, its mathematical representation also has 
undergone the discontinuous change, and we speak of a `quantum jump'."  

``A real difficulty in understanding the interpretation occurs when one asks 
the famous question: But what happens `really' in an atomic event?"  

``If we want to describe what happens in an atomic event, we have to realize 
that the word `happens' can apply only to the observation, not to the state
of affairs between the two observations.  It [ the word `happens' ]
applies to the physical, not the psychical act of observation, and we
may say that the transition from the `possible' to the `actual' takes
place as soon as the interaction of the object with the measuring
device, and therefore with the rest of the world, has come into play;
it is not connected with the act of registration of the result in the
mind of the observer.  The discontinuous change in the probability
function, however, occurs with the act of registration, because it is
the discontinuous change in our knowledge in the instant of
recognition that has its image in the discontinuous change in the
probability function."

This explanation uses two distinct modes of description. One is a pragmatic
knowledge-based description in terms of the Copenhagen concept of the 
discontinuous change of the quantum-theoretic probability function at the 
registration of new knowledge in the mind of the observer. The other is 
an ontological description in terms of `possible' and `actual',
and `interaction of object with the measuring device'.  The latter 
description is an informal supplement to the strict Copenhagen interpretation.
I say `informal supplement' because this ontological part is not tied 
into quantum theoretical formalism in any precise way. It assuages the
physicists' desire for an intuitive understanding of what could be going on 
behind the scenes, without actually interfering with the workings of the 
pragmatic set of rules.

Heisenberg's transition from `the possible' to `the actual' 
at the dumb measuring device was shown to be a superfluous and
needless complication by von Neumann's analysis 
of the quantum process of measurement (von Neumann, 1932, 
Chapter VI). I shall discuss that work later, but note here 
only the key conclusion. von Neumann introduced the 
measuring instruments and the body/brains of the community 
of human observers into the quantum state, which is
quantum theory's only representation of ``physical reality''. 
He then showed that if an observer experiences the fact that, for 
example, `the pointer on a measuring device has swung to the right', 
then this increment in the observer's knowledge can be 
associated exclusively with a reduction (i.e., sudden change)
of the state of the brain of that observer to the part of that 
brain state that is compatible with his new knowledge. No 
change or reduction of the quantum state at the dumb measuring
device is needed: no change in ``knowledge'' occurs there. 
This natural association of human 
``knowings'' with  events in human brains allows the 
`rules' of the Copenhagen interpretation pertaining to 
``our knowledge'' to be represented in a natural ontological 
framework. Indeed, any reduction event at the measuring 
device itself would, strictly speaking, disrupt in principle 
the validity of the predictions of quantum theory. Thus the
only natural ontological place to put the reduction associated 
with the increases in knowledge upon which the Copenhagen 
interpretation is built is in the brain of the person 
whose knowledge is increased. 

My purpose in what follows is to reconcile the insight of 
the founders of quantum theory, namely that the mathematical 
formalism of quantum theory is about our knowledge, with the 
demand of Einstein that basic physical theory be about 
nature herself. I shall achieve this reconciliation 
by incorporating human beings, including both their 
body/brains and their conscious experiences, into the quantum 
mechanical description of nature. 

The underlying commitment here is to the basic quantum
principle that information is the currency of reality, not matter: 
the universe is an informational structure, not a substantive one. 
This fact is becoming ever more clear in the empirical studies of the 
validity of the concepts of quantum theory in the context of
complex experiments with simple combinations of correlated
quantum systems, and in the related development of quantum 
information processing. Information-based language works 
beautifully, but substance-based language does not work at all..

\hspace{.2in}
\begin{center}
{\bf        Mind/Brain Dynamics: Why Quantum Theory Is Needed    }
\end{center}
\vspace{.1in}

A first question confronting a classically biased mind-brain researcher 
is this: How can two things so differently described and conceived as 
substantive matter and conscious thoughts interact in any rationally 
controlled and scientifically acceptable way. Within the classical framework 
this is impossible. Thus the usual tack has been to abandon or modify 
the classical conception of mind while clinging tenaciously to the 
``scientifically established'' classical idea of matter,
even in the face of knowledge that the classical idea of matter is 
now known by scientists to be profoundly and fundamentally mistaken, and  
mistaken not only on the microscopic scale, but on the scale of 
meters and kilometers as well (Tittel, 1998). Experiments show that
our experiences of instruments  cannot possibly be just the 
passive witnessing of macroscopic physical realities that exist and behave 
in the way that the ideas of classical physical theory say that macroscopic 
physical realities ought to exist and behave. 

Scientists and philosophers intent on clinging to familiar classical 
concepts normally argue at this point that whereas long-range quantum 
effects can be exhibited under rigorous conditions of isolation and 
control, all quantum effects will be wiped out in warm wet brains on a 
very small scale, and hence classical concepts will be completely adequate 
to deal with the question of the relationship between our conscious 
thoughts and the large-scale brain activities with which they are almost 
certainly associated. 
 
That argument is incorrect. The emergence of classical-type relationships 
arise from interactions between a system and its environment. These 
interactions induce correlations between this system and its environment
that make certain typical quantum interference effects difficult to observe 
{\it in practice}, and that allow certain practical computations to be  
simplified by substituting a classical system for a quantum one.
However, these correlation (decoherence) effects definitely do not entail 
the true emergence --- even approximately --- of a single classically 
describable system. (Zurek, 1986, p.89 and Joos, 1986, p.12). 
In particular, if the subsystem of interest is a brain  then interactions 
between its parts produce a gigantic jumble of partially interfering 
classical-type states: no single approximately classical reality emerges. 
Yet if no --- even-approximate --- single classical reality emerges at any 
macroscopic scale, but only a jumble of partially interfering quantum states, 
then the investigation of an issue as basic as the nature of the mind-brain
connection ought {\it in principle} to be pursued within an exact framework,
rather than crippling the investigation from the outset by replacing correct 
principles by concepts known to be fundamentally and grossly false, just 
because they allow certain {\it practical} computations to be simplified.

This general argument is augmented by a more detailed 
examination of the present case. The usual argument for the 
approximate {\it pragmatic} validity of a classical 
conceptualization of a system is based on assumptions about 
the nature of the question that is put to nature. The 
assumption in the usual case is that this question 
will be about something like the position of a visible object.
Then one has a clear separation of the world into its 
pertinent parts: the unobservable atomic subsystem, the 
observable features of the instrument, and unobserved 
features of the environment, including unobserved 
micro-features of the instrument. 
The empirical question is about the observable features of 
the instrument. These features are essentially just the 
overall position and orientation of a visible object. 

But the central issue in the present context is precisely 
the character of the  brain states that are associated with 
conscious experiences. It is not known a priori whether
or how a self-observing quantum system separates into these 
various parts. It is not clear, a priori, that a 
self-observing brain can be separated into components
analogous to observer, observee, and environment. 
Consequently,  one cannot rationally impose prejudicial 
assumptions --- 
based on pragmatic utility in simple cases in which the 
quantum system and measuring instrument are two distinct 
systems both external to the human observer, and strongly 
coupled to an unobservable environment --- in this vastly 
different present case, in which the quantum system being 
measured, the observing instrument, and 
``the observer'' are aspects of one unified body/brain/mind
system observing itself.

In short, the practical utility of classical concepts in 
certain special situations arises from the very special 
forms of the empirical questions that are to be asked in 
those situations. Consequently, one must revert to the basic
physical principles in this case where the special 
conditions of separation fail, and the nature of the 
questions put to nature can therefore be quite different.

The issue here is not whether distinct objects that we observe via our senses
can be treated as classical objects. It is whether in the description of 
the complex inner workings of a thinking human brain it is justifiable to 
assume --- not just for certain simple practical purposes, but as a matter 
of principle --- that this brain is made up of tiny interacting parts of 
a kind known not to exist. 

The only rational scientific way to proceed in this 
case of a mind/brain observing itself is to start from basic 
quantum theory, not from a theory that is known to be
profoundly incorrect.

The vonNeumann/Wigner ``orthodox'' quantum formalism that I employ 
automatically and neatly encompasses all quantum and classical predictions, 
including the transition domains between them. It automatically incorporates 
all decoherence effects, and the partial ``classicalization'' 
effects that they engender.

\vspace{.2in}
\begin{center}
{\bf              vonNeumann/Wigner Quantum Theory       }
\end{center}
\vspace{.1in}

Wigner used the word ``orthodox'' to describe the formulation of quantum 
theory developed by von Neumann. It can be regarded as a partial
ontologicalization of its predecessor, Copenhagen quantum theory. 

The central concept of the Copenhagen interpretation of quantum theory,
as  set forth by the founders at the seminal Solvay conference of
1927, is that the basic mathematical entity of the theory, the quantum state
of a system, represents ``our knowledge'' of the system, and the reduced
state represents our more precise knowledge after measurement. 

In the strict Copenhagen view, the quantum state
is always the state of a limited system that does not
include the instruments that we use to {\it prepare} that system or later
to {\it measure} it. Our relevant experiences are those that we 
described as being our observations of the observable features of these 
instruments. 

To use the theory one needs relationships between the mathematical quantities 
of the theory and linguistic specifications on the observable features of the 
instruments. These specifications are couched in the  language that we
use to communicate to our technically trained associates what we have done
(how we have constructed our instruments, and put them in place) and what we 
have learned (which outcomes have appeared to us). Thus pragmatic quantum 
theory makes sense only when regarded as a part of a larger enveloping 
language that allows us describe to each other the dispositions of the 
instruments and ordinary objects that are relevant to the application we 
make. The connections between these linguistic specifications and the 
mathematical quantities of the theory are fixed, fundamentally, by the 
empirical calibrations of our instruments. 

These calibration procedures do not, however, fully exploit all that we 
know about the atomic properties of the instruments. 

That Bohr was sensitive to this deficiency, is shown by following passage:

``On closer consideration, the present formulation of quantum mechanics,
in spite of its great fruitfulness, would yet seem no more than a first
step in the necessary generalization of the classical mode of description,
justified only by the possibility of disregarding in its domain of application
the atomic structure of the measuring instruments. For a correlation of still
deeper lying laws of nature ... this last assumption can no longer be 
maintained and we must be prepared for a ... still more radical renunciation
of the usual claims of so-called visualization. (Bohr, 1936, p,293-4)''

Bohr was aware of the work in this direction by John von Neumann (1932), but
believed von Neumann to be on a wrong track. Yet the opinion of 
many other physicists is that von Neumann made the right moves: he brought 
first the measuring instruments, and eventually the entire physical 
universe, including the human observers themselves, into the physical system 
represented by the quantum state. The mathematical theory allows one to
do this, and it is unnatural and problematic to do otherwise: any other
choice would be an artifact, and would create problems associated
with an artificial separation of the unified physical system into differently 
described parts. This von Neumann approach, in contrast to the Copenhagen 
approach, allows the quantum theory to be applied both to cosmological 
problems, and to the mind-body problem.
 
Most efforts to improve upon the original Copenhagen quantum theory
are based on von Neumann's formulation. That includes the present work. 
However, almost every other effort to modify the Copenhagen formulation aims 
to improve it by {\it removing} the consciousness of the observer from quantum 
theory: they seek to bring quantum theory in line with the basic philosophy 
of the superceded classical theory, in which consciousness is 
imagined to be a disconnected passive witness. 

I see no rationale for this retrograde move. Why should we impose on our
understanding of nature the condition that consciousness not be an integral
part of it, or an unrealistic stricture of impotence that is belied by the 
deepest testimony of human experience, and is justified only by a theory 
now known to be fundamentally false, when the natural form of the 
superceding theory makes experience efficacious?

I follow, therefore, the von Neumann/Wigner [vN/W] formulation, in which 
the entire physical world is represented by a quantum mechanical state, 
and each thinking human being is recognized as an aspect of the total reality: 
each thinking human being is a body/brain/mind system, consisting of a 
sequence of conscious events, called knowings, bound together by the 
physical structure that is his body/brain.

However, the basic idea, and the basic rules, of  Copenhagen quantum theory 
are strictly maintained: the quantum state continues to represent knowledge, 
and each experiential increment in knowledge, or knowing, is accompanied by a 
reduction of the quantum state to a form compatible with that increase in 
knowledge. 

By keeping these connections intact one retains both the close 
pragmatic link between the theory and empirical knowledge, which is 
entailed by the quantum rules, and also the dynamical efficacy of 
conscious experiences, which follows from the action of the `reduction of 
the quantum state' that, according to the quantum rules, is the image in 
the physical world of the conscious event. 

In this theory, each conscious event has as its physical image not a 
reduction of the state of some small physical system that is external 
to the body/brain of the person to whom the experience belongs, as specified 
by the Copenhagen approach. Rather, the reduction is in that part of the 
state of the universe that constitutes the state of the body/brain of the 
person to whom the experience belongs: the reduction actualizes the pattern 
of activity that is sometimes called the ``neural correlate'' of that 
conscious experience. The theory thus ties in a practical way into the 
vast field of mind-brain research: i.e., into studies
of the correlations between, on the one hand, brain 
activities of a subject, as measured by instrumental probes and described
in physical terms, and, on the other hand, the subjective experiences, 
as reported by the subject, and described in the language of 
``folk psychology'' [i.e., in terms of 
feelings, beliefs, desires, perceptions, and the other
psychological features.]

My aim now is to show in more detail how the conscious intentions of a human 
being can influence the activities of his brain. To do this I must first 
explain the two important roles of the quantum observer.

\vspace{.2in}
\begin{center}
{\bf              The Two Roles of the Quantum Observer          }
\end{center}
\vspace{.1in}

Most readers will have heard of the Schroedinger equation: it is the 
quantum analog of Newton's and Maxwell's equations 
of motion of classical mechanics. The Schroedinger equation,
like Newton's and Maxwell's equations, is deterministic: given the motion of 
the quantum state for all times prior to the present, the motion for all 
future time is fixed, insofar as the Schroedinger equation is satisfied for 
all times. 

However, the Schroedinger equation fails when an increment of knowledge 
occurs: then there is a sudden jump to a `reduced' state, which represents the 
new state of knowledge. This jump involves the well-known element of
quantum randomness. 

A superficial understanding of quantum theory might easily lead one to conclude
that the entire dynamics is controlled by just the combination of the
local-deterministic Schroedinger equation and the elements of quantum
randomness. If that were true then our conscious experiences would again 
become epiphenomenal side-shows.

To see beyond this superficial appearance one must look more closely at
the two roles of the observer in quantum theory. 

Niels Bohr (1951, p.223), in recounting the important events at the 
Solvay Conference of 1927, says:

``On that occasion an interesting discussion arose also about how to speak 
of the appearance of phenomena for which only predictions of a statistical 
nature can be made. The question was whether, as regards the occurrence of
individual events, we should adopt the terminology proposed by Dirac,
that we have to do with a choice on the part of `nature' or, as suggested 
by Heisenberg, we should say that we have to do with a choice on the part
of the `observer' constructing the measuring instruments and reading their
recording.''

Bohr stressed this choice on part of the observer:

``...our possibility of handling the measuring instruments allow us only
to make a choice between the different complementary types of phenomena
we want to study.''

The observer in quantum theory does more than just read the recordings. 
He also chooses which question will be put to Nature: which aspect of nature 
his inquiry will probe. I call this important function of the observer 
`The Heisenberg Choice', to contrast it with the `Dirac Choice', which is 
the random choice on the part of Nature that Dirac emphasized. 

According to quantum theory, the Dirac Choice is a choice between alternatives
that are specified by the Heisenberg Choice: the observer must first specify 
what aspect of the system he intends to measure or probe, and then put in 
place an instrument that will probe that aspect. 

In  quantum theory it is the observer who both poses the question, and 
recognizes the answer. Without some way of specifying what the question is, 
the quantum rules will not work: the quantum process grinds to a halt.

Nature does not answer, willy-nilly, all questions: it answers only properly 
posed questions.

A question put to Nature must be one with a Yes-or-No answer, or a sequence 
of such questions. The question is never of the form ``Where will object O 
turn out to be?'',  where the possibilities range in a smooth way over a 
continuum of values. The question is rather of a form such as: ``Will the 
center of object O --- perhaps the pointer on some instrument --- be found by 
the observer to lie in the interval between 6 and 7 on some specified `dial'?''

The human observer poses such a question, which must be such that the 
answer Yes is experientially recognizable. Nature then delivers the answer, 
Yes or No. Nature's answers are asserted by quantum theory to conform to 
certain statistical conditions, which are determined jointly by the question 
posed and the form of the prior state (of the body/brain of the 
observer.) The observer can  examine the answers that Nature gives, in a 
long sequence of trials with similar initial conditions, and check 
the statistical prediction of the theory.

This all works well at the pragmatic Copenhagen level, where the observer
stands outside the quantum system, and is simply accepted for what he
empirically is and does. But what happens when we pass to the vN/W ontology?
The observer then no longer stands outside the quantum system: he becomes a 
dynamical body/brain/mind system that is an integral dynamical part of the 
quantum universe. 

The basic problem that originally forced the founders of quantum theory 
to bring the human observers into the theory was that the evolution of the
state via the Schroedinger equation does not fix or specify where and when
the question is posed, or what the question actually is. This problem
was resolved by placing this issue in the hands and mind of the external 
human observer. 

Putting the observer inside the system does not, by itself, resolve this 
basic problem: the Schroedinger evolution alone remains unable to specify 
what the question is. Indeed,  this bringing of the human observer into 
the quantum system intensifies the problem, because there is no longer 
the option of shifting the problem away, to some outside agent. Rather, the 
problem is brought to a head, because the human agent is precisely the 
quantum system that is under investigation.

In the Copenhagen formulation the Heisenberg choice was made by the mind 
of the external human observer. I call this process of choosing the 
question the Heisenberg process. In the vN/W formulation this choice 
is not made by the local deterministic Schroedinger process and the 
global stochastic Dirac process. So there is still an essential need
for a third process, the Heisenberg process. Thus the agent's mind can 
continue to play its key role. But the mind of the human agent is now an 
integral part of the dynamical body/brain/mind. We therefore have, now, 
an intrinsically more complex dynamical situation, one in which a 
person's conscious thoughts can --- and evidently must, if no new 
element is brought in, --- play a role that is not reducible to the 
combination of the Schroedinger and Dirac processes. In an evolving
human brain governed by ionic concentrations and electric-magnetic 
field gradients, and other continuous field-like properties, rather than 
sharply defined properties, or discrete well-defined ``branches'' of 
the wave function, the problem of specifying, within this amorphous and 
diffusive context, the well-defined question that is put to nature is 
quite nontrivial.

Having thus identified this logical opening for efficacious human mental 
action, I now proceed to fill in the details of how it might work.

\vspace{.2in}
\begin{center}
{\bf         How Conscious Thoughts Could Influence Brain Process   }
\end{center}
\vspace{.1in}

Information is the currency of reality. That is the basic message of 
quantum theory.

The basic unit of information is the ``bit'': the answer `Yes' or `No'
to some specific question.

In quantum theory the answer `Yes' to a posed question is associated
with an operator $P$ that depends on the question. The defining property
of a projection operator is that $P$ squared equals $P$: asking the 
very same question twice it is the same as asking it once. The 
operator associated with the answer `No' to this same question is $1-P$.
Note that $(1-P)$ is also a projection operator: 
$(1-P)^2 = 1 -2P + P^2 = 1-2P +P=(1-P)$. 

To understand the meaning of these operators $P$ and $(1-P)$ it is 
helpful to imagine a trivial classical example. Suppose a motionless 
classical heavy point-like particle is known to be in a box that is 
otherwise empty. Suppose a certain probability function F represents all 
that you know about the location of this particle. Suppose you then send
some light through the left half of the box that will detect the particle
if it is in the left half of the box, but not tell you anything about
where in the left half of the box the particle lies. Suppose, moreover,
that the position of the particle is undisturbed by this observation.  
Then let P be the operator that acting on any function $f$ sets that 
function to zero in the right half of the box, but leaves it unchanged 
in the left half of the box. Note that two applications of P has exactly
the same effect as one application, $P^2 = P$. The question put to nature
by your probing experiment is: ``Do you now know that particle is in the 
left half of the box? Then the function PF represents, apart from an overall 
normalization factor, your new state of knowledge if the answer to the 
posed question was YES. Likewise, the function (1-P)F represents, apart 
from overall normalization, the new probability function, if the answer 
was NO.

The quantum counterpart of F is the operator S. Operators are like
functions that do not commute: the order in which you apply them matters. 
The analog of $PF \equiv PFP$ is $PSP$, and the analog of 
$(1-P)F \equiv (1-P)F(1-P)$ is $(1-P)S(1-P)$. 

This is how the quantum state represents information and knowledge,
and how increments in knowledge affect the quantum state.

I have described in my book (Stapp, 1993, Ch 6) my conception of how the
quantum mind/brain works. It rests on some ideas/findings of William James.

William James(1910, p.1062) says that:

 ``a discrete composition is what actually obtains
in our perceptual experience. We either perceive nothing, or something that 
is there in sensible amount. This fact is what in psychology is known as the
law of the `threshold'. Either your experience is of no content, of no change,
or it is of a perceptual amount of content or change. Your acquaintance with
reality grows literally by buds or drops of perception. Intellectually and on 
reflection you can divide these into components, but as immediately given they
come totally or not at all.'' 

This wholeness of each perceptual experience is a main conclusion, and theme,
of Jamesian psychology. It fits neatly with the quantum ontology.

Given a well posed question about the world to which one's attention is 
directed quantum theory says that nature either gives the affirmative 
answer, in which case there occurs an experience describable as ``Yes, I 
perceive it!'' or, alternatively,  no experience occurs in connection with 
that question.

In  vN/W theory the `Yes' answer is represented by a projection
operator P that acts on the degrees of freedom of the brain of the observer, 
and reduces the state of this brain --- and also the state S of the universe
--- to one compatible with that answer `Yes': S is reduced to PSP. If the 
answer is `No', then the projection operator $(1-P)$ is applied to the 
state S: S is reduced to (1-P)S(1-P). [See Stapp (1998b) for technical 
details.]

James (1890, p.257) asserts that each conscious experience, though it comes 
to us whole, has a sequence of temporal components ordered in accordance 
with the ordering in which they have entered into one's stream  of 
conscious experiences. These components are like the columns in a marching 
band: at each viewing only a subset of the columns is in front of the viewing 
stand. At a later viewing a new column has appeared on one end, and one has
disappeared at the other. (cf. Stapp, 1993, p. 158.) It is this possibility 
of having a sequence of different components present in a single thought that 
allows conscious analysis and comparisons to be made.

Infants soon grasp the concept of their bodies in interaction with a world
of persisting objects about them. This suggests that the brain of an
alert person normally contains a ``neural'' representation of the current 
state of his body and the world about him. I assume that such a representation
exists, and call it the body-world schema. (Stapp, 1993, Ch. 6)

Consciously directed action is achieved, according to this theory,
by means of a `projected' (into the future) temporal component of the 
thought, and of the body-world schema actualized by the thought: the 
intended action is represented in this projected component as a mental image 
of the intended action, and as a corresponding representation in the brain, 
(i.e., in a body-world schema)  of that intended action. The neural activities
that automatically flow from the associated body-world schema tend to bring 
the intended bodily action into being.

The coherence and directedness of a person's  stream of consciousness
is maintained, according to this theory, because the instructions effectively
issued to the unconscious processes of the brain by the natural dynamical
unfolding that issues from the actualized body-world schema include not only 
the instructions for the initiation or continuation of motor actions but also 
instructions for the initiation or continuation of mental processing. This 
means that the actualization associated with one thought leads physically to 
the emergence of the propensities for the occurrence of the next thought, 
or of later thoughts. (Stapp, 1993, Ch. 6) 

The idea here is that the action --- on the state $S$ --- of the projection 
operator P that is associated with a thought $T$ will actualize a pattern of 
brain activity that will dynamically evolve in such a way as to tend to 
create a subsequent state that is likely to achieve the intention of the 
thought $T$. The natural cause of this  positive correlation between 
the experiential intention of the thought $T$ and the matching confirmatory
experience of a succeeding thought $T'$ is presumably set in place during the 
formation of brain structure, in the course of the person's  interaction
with his environment, by the reinforcement of brain structures that result in
empirically successful pairings between experienced intentions and subsequently
experienced perceptions. These can be physically compared because both are 
expressed physically by similar body-world schemas.

As noted previously,
the patterns of brain activity that are actualized by an event unfold not 
only into instructions to the motor cortex to institute intended
motor actions. They unfold also into instructions for the creation of the 
conditions for the next experiential event. But the Heisenberg uncertainties
in, for example, the locations of the atomic and ionic constituents of the 
nerve terminals, and more generally of the entire brain, necessarily engender 
a quantum diffusion in the evolving state of the brain. Thus the dynamically 
generated state that is the pre-condition for the next event will not 
correspond exactly to a well defined unique question: some `scatter' will
invariably creep in. However, a specific question must be posed in order for 
the next quantum event to occur!

This problem of how to specify ``the next question'' is the central problem
in most attempts to `improve' the Copenhagen interpretation by 
excluding ``the observer''. If one eliminates the observer, then something
else must be brought in to fix the next question: i.e., to make the Heisenberg 
choice.

The main idea here is to continue to allow the question to be posed by the 
`observer', who is now an integral part of the quantum system: the observer is
a body/brain/mind subsystem. The Heisenberg Choice, which is the choice
of an operator P that acts macroscopically, as a unit, on the observing 
system, is not fixed by the Schroedinger equation, or by the Dirac Choice, 
so it is most naturally fixed by the experiential part of that system, which 
seems to pertain to  macroscopic aspects of brain activity taken as units.

Each experience is asserted to have
an intentional aspect, which is its experiential goal or aim, and an 
attentional aspect, which is an experiential focussing on an updating of 
the current status of the person's idea of his body, mind, and environment.

When an action is initiated by some thought, part of the instruction
is normally to monitor, by attention, the ensuing action, in order to check 
it against the intended action.

In order for the appropriate experiential check to occur, {\it the appropriate 
question must be asked.} The intended action is formulated in experiential 
terms, and the appropriate monitoring question is whether this intended 
experience matches the subsequently occurring experience. {\it This 
connection has the form of the transference of an experience defined by 
the intentional aspect of an earlier experience into the experiential 
question attended to --- i.e., posed --- by a later experience.}

This way of closing the causal gap associated with the Heisenberg Choice
introduces two parallel lines of causal connection in the body/brain/mind 
system. On the one hand, there is the physical line that unfolds --- under the
control of the local deterministic Schroedinger equation --- from a prior 
event, and that generates the physical {\it potentialities} for succeeding 
possible events. Acting in parallel to this physical line of causation, there 
is a mental line of causation that transfers the experiential intention of 
an earlier event into an experiential attention of a later event. These two 
causal strands, one physical and one mental, join to form 
the physical and mental poles of a succeeding quantum event.

In this model there are three intertwined factors in the causal structure:
(1), the local causal structure generated by the Schroedinger equation; 
(2), the Heisenberg Choices, which is based on the experiential aspects of 
the body/brain/mind subsystem that constitutes a person; and
(3), the Dirac Choices on the part of nature.

The point of all this is that there is within the vN/W ontology a logical 
necessity, in order for the quantum process to proceed,  for {\it some 
process} to fix the Heisenberg Choice of the operator $P$, which acts
over an extended portion of the body/brain of the person. Neither the 
Schroedinger evolution nor the Dirac stochastic choice can do the job. The
only other known aspect of the system is our conscious experience. It is 
possible, and natural, to use this mind part of body/brain/mind system 
to produce the needed choice.

The mere logical possibility of a mind-matter interaction such as this,
within the vN/W formulation, indicates that quantum theory has the 
potential of permitting the experiential aspects of reality to enter into 
the causal structure of body/brain/mind dynamics, and to enter in a way 
that is not fully reducible to a combination of local mechanical causation 
specified by the Schroedinger equation and the random quantum choices.
The requirements of quantum dynamics {\it demand} some further process,
and an experienced-based process that fits both our ideas about our
psychological make up and also the quantum rules that connect our experiences
to the informational structure carried by the evolving physical state
of the brain seems to be the perfect candidate.

What has been achieved here is, of course, just a working out in more 
detail of Wigner's idea that quantum theory, in the von Neumann form,
allows for mind --- pure conscious experience --- to {\it interact} with 
the `physical' aspect of nature, as that aspect is represented in quantum 
theory. What permits this interaction is the fact that the physical aspect 
of nature, as it is represented in quantum theory, is informational in 
character, and hence links naturally to increments in knowledge. 
Because each increment in knowledge acts directly upon the quantum state,
and reduces it to the informational structure compatible with the new
knowledge, there is, right from the outset, an action of mind on the 
physical world. I have just worked out a possible
scenario in more detail, and in particular have emphasized how the
causal gap associated with the Heisenberg Choice allows mind to enter
into the dynamics in a way that is quite in line with our intuition
about the efficacy of our thoughts. It is therefore simply wrong
to proclaim that the findings of science entail that our intuitions 
about the nature of our thoughts are necessarily illusory or false. Rather, 
it is completely in line with contemporary science to hold  our thoughts 
to be causally efficacious, and reducible neither to the local deterministic 
Schroedinger process, nor to that process combined with stochastic
Dirac choices on the part of nature.

\begin{center}
{\bf Idealism,  Materialism, and Quantum Informationism.}
\end{center}
\vspace{.1in}

I have stressed just now the idea-like character of the physical state
of the universe, within vN/W quantum theory. This suggests that the theory
may conform to the tenets of idealism. This is partially true. The quantum
state undergoes, when a fact become fixed in a local region, a sudden jump
that extends over vast reaches of space. This gives the physical state the
character of a representation of knowledge rather than a representation
of substantive matter. When not jumping the state represents
potentialities or probabilities for actual events to occur.
Potentialities and probabilities are normally conceived to be idea-like
qualities, not material realities. So as regards the intuitive conception
of the intrinsic nature of {\it what is represented} within the theory
by the physical state it certainly is correct to say that it is idea-like.

On the other hand, the physical state has a mathematical structure,
and a behaviour that is governed by the mathematical properies.
It evolves much of the time in accordance with local deterministic laws
that are direct quantum counterparts of the local deterministic laws of
classical mechanics. Thus as regards various structural and causal
properties the physical state certainly has aspects that we normally
associate with matter.

So this vN/W quantum conception of nature ends up having both idea-like and
matter-like qualities. The causal law involves two complementary modes of
evolution that, at least at the present level theoretical development, are
quite distinct. One of these modes involves a gradual change that is governed
by local deterministic laws, and hence is matter-like in character. The other
mode is abrupt, and is idea-like in {\it two} respects.

This hybrid ontology can be called an information-based reality.  Each
answer, Yes or No, to a quantum question is one bit of information that is
generated by a mental-type event. The physical repository of this information
is the quantum state of the universe: the new information is recorded as 
a reduction of the quantum state of the universe to a new form, which then 
evolves deterministically in accordance with the Schroedinger equation. Thus, 
according to this quantum conception of nature, the physical universe
 --- represented by the quantum state --- is a repository of evolving 
information that has the dispositional power to create more information.

This hybrid ontology can be called an information-based reality.  Each
answer. Yes or No, to a quantum question is one bit of information that is
generated by a mental-type event. This event is registered as a reduction
of the quantum state of the universe to a new form. This information is stored in 
this state, which evolves deterministically in accordance  
with the Schroedinger equation. Thus, according to the quantum
conception, the physical universe --- represented by the 
quantum state --- is a repository evolving information that has
the dispositional power to create more information.

\vspace{.2in}
\begin{center}
{\bf             Quantum Zeno Effect and The Efficacy of Mind       }
\end{center}
\vspace{.1in}

In the model described above the specifically mental effects are
expressed solely through the choice and the timings of the questions posed. 
The question then arises as to whether just the choices about which
questions are asked, with no control over which answers are returned,
can influence the dynamical evolution of a system.

The answer is `Yes':  the evolution of a quantum state can be greatly 
influenced by the choices and timings of the questions put to nature.

The most striking example of this is the Quantum Zeno Effect. 
(Chui, Sudarshan, and Misra, 1977, and Itano, et al. 1990). In quantum theory
if one poses repeatedly, in very rapid succession, the same Yes-or-No
question, and the answer to the first of these posings is Yes, then in the 
limit of very rapid-fire posings the evolution will be confined to the
subspace in which the answer is Yes: the effective Hamiltonian will change
from H to PHP, where P is the projection operator onto the Yes states.
This means that evolution of the system is effectively ``boxed in''
in the subspace where the answer continues to be Yes, if the question is posed
sufficiently rapidly, even if it would otherwise run away from that region.

This fact that the Hamiltonian is effectively changed in this macroscopic way 
shows that the choices and timings of which questions are asked can affect 
observable properties.

\vspace{.2in}
\begin{center}
{\bf                       Free Will and Causation            }
\end{center}
\vspace{.1in}

Personal responsibility is not reconciled with the quantum understanding of
causation by making our thoughts {\it free}, in the sense of being 
completely unconstrained by anything at all. It is solved, 
rather, by making our thoughts  {\it part} of the causal structure of the 
body/brain/mind system, but a part that is not under the complete dominion 
of myopic ({\it i.e., microlocal}) causation and random chance. Our thoughts 
then become aspects of the causal structure that are {\it entwined} with 
the micro-physical and random elements, yet are not completely reducible to 
them, or replaceable by them.

\vspace{.2in}
\begin{center}
{\bf              Pragmatic Theory of the Mind/Brain          }
\end{center}
\vspace{.1in}

This vN/W theory gives a conceivable ontology. However, for practical 
purposes it can be viewed as a pragmatic theory of the human psycho-physical 
structure. It is deeper and more realistic than the Copenhagen version 
because it links our thoughts not directly to objects (instruments) in the 
external world, but rather to patterns of brain activity. It provides a 
theoretical structure based explicitly on the two kinds of data at our 
disposal, namely the experiences of the subject, as he describes these 
experiences to himself and his colleagues, and the experiences of the 
observers of that subject, as they describe their experiences to themselves 
and their colleagues. These two kinds of descriptions are linked together 
by a theoretical structure that neatly, precisely, and automatically accounts,
in a single uniform and practical way, for all known quantum and classical 
effects. But, in contrast to the classical-physics based model, it has a 
ready-made place for an efficacious mind, and provides a rational  
understanding of how such a mind could be causally enmeshed with brain
processes.

If one adopts this pragmatic view then one need never consider 
the question of nonhuman minds: the theory then covers, by definition, 
the science that we human beings create to account 
for the structure of our human experiences. 

This pragmatic theory should provide satisfactory basis for
a rational science of the human mind/brain. It gives a structure that 
coherently combines the psychological and physical aspect of
human behavior. However, it cannot be expected to be exactly true, for it 
would entail the existence of collapse events associated with increments in 
human knowledge, but no analogous events associated with non-humans. 

One cannot expect our species to play such a special role in nature.
So this human-based pragmatic version must be understood, from the
ontological standpoint, as merely the first stage in the development of 
a better ontological theory: one that accommodates the evolutionary
precursors to the human knowings that the pragmatic theory is 
based upon. 

So far there is no known empirical evidence for the existence of any
reduction events not associated with human knowings. This impedes,
naturally, the development of a science that encompasses such other
events.

\vspace{.2in}
\begin{center}
{\bf       Future Developments: Representation and Replication        }
\end{center}
\vspace{.1in}

The primary purpose of this paper has been to describe the general features 
of a pragmatic theory of the human mind/brain that allows our thoughts to be 
causally efficacious yet not controlled by local-mechanistic laws combined 
with random chance. Eventually, however, one would like to expand this 
pragmatic version into a satisfactory ontology theory. 

Human experiences are closely connected to human brains. Hence 
events similar to human experiences would presumably not exist either in 
primitive life forms, or before life began. Hence a more general theory that 
could deal with the {\it evolution of consciousness} would presumably have to 
be based on something other than the ``experiential increments in  knowledge'' 
that were the basis of the pragmatic version described above.

Dennett (1994, p.236) identified intentionality (aboutness) as a phenomenon 
more fundamental than consciousness, upon which he would build his theory
of consciousness. `Aboutness' pertains to representation:
the representation of one thing in another. 

The body-world schema is the brain's representation of the body and its
environment. Thus it constitutes, in the theory of consciousness described 
above, an element of ``aboutness'' that could be seized upon as the basis 
of a more general theory.

However, there lies at the base of the quantum model described above an 
even more rudimentary element: self-replication. The basic process in the
model is the creation of events that create likenesses of themselves.  
This tendency of thoughts to create likenesses of themselves,
helps to keep a train of thought on track.

Abstracting from our specific model of human consciousness one sees
the skeleton of a general process of self-replication. 

Fundamentally, the theory described above is a theory of events, where
each event has an {\it attentional} aspect and an {\it intentional} aspect.
The attentional aspect of an event specifies an item of
information that fixes the operator $P$ associated with that event.
The intentional aspect of the event specifies the functional property injected
into the dynamics by the action of $P$ on $S$. This functional property is a
tendency of the Schroedinger-directed dynamics to produce a future event 
whose attentional aspect is the same as that of the event that is producing
this tendency. The effect of these interlocking processes is to inject 
into the dynamics a directional tendency, based on approximate 
self-replication, that acts against the chaotic diffusive tendency 
generated by the Schroedinger equation. Such a  process could occur  
before the advent of our species, and of life itself, and it could
contribute to their emergence.

\vspace{.2in}
\begin{center}
{\bf                              Conflation and Identity         }
\end{center}
\vspace{.1in}

A person's thoughts and ideas appear --- to that person himself --- to be able
to do things:  a person's mental states seem to be able cause his body to
move
about in intended ways. Thus thoughts  seem to have functional power. Indeed,
the idea of {\it functionalism} is that what makes thoughts and other mental
states what they  are is precisely their functional power: e.g., my pain
is a pain by virtue of its functional or causal relationship
to other aspects of the body/brain/mind system. Of course, this would be
merely
a formal definition of the term ``mental state'' if it did not correspond
to the occurrence of an associated element in a person's stream of
consciousness: in the context of the present study --- of the connection
between our brains and our inner experiential lives --- the occurrence of a
mental state in a person's mind  is supposed to mean the occurrence of a
corresponding element in his stream of consciousness.

The identity theory of mind claims that each mental state is {\it identical}
to some process in a brain. But combining this idea with the
classical-physics
conception of the physical universe leads to problems.
They stem from the fact that the precepts of classical physical
theory entail that the entire causal structure of any complex physical system
is completely determined by its microscopic physical structure alone.
Alternative high-level descriptions of certain complex physical systems
might be far more useful to us in practice, but they are in principle
redundant and unnecessary if the principles of classical physics hold. Thus
it is accurate to say that the heat of the flame caused the paper to
ignite, or that the tornado ripped the roofs off of the houses and left a
path of destruction. But according to the precepts of classical physical
theory the high-level causes are mere mathematical reorganizations of
microscopic causes that are completely explainable micro-locally
within classical physical theory. Nothing is needed beyond mathematical
reorganization and --- in order for us to be able to apply the theory --- 
the assumption that we can empirically know, through observations via our 
senses,the approximate relative locations and shapes of sufficiently large
macroscopically localized assemblies of the microscopic physical elements
that the theory posits.

In the examples just described our experiences themselves are not the causes
of the ignition or destruction: our experiences merely help us to identify
the causes. In fact, the idea behind classical physical theory is that the
local physical variables of the theory represent a collection of
ontologically distinct physical realities each of whose ontological status is
(1), intrinsically microlocal, (2), ontologically independent of our
experiences, and (3), dynamically non-dependent upon experiences. That is why
quantum theory was such a radical break with tradition: in quantum theory the
physical description became enmeshed with our experiential knowledge, and the
physical state became causally dependent upon our mental states.

Quantum theory is, in this respect, somewhat similar to the identity
theory of mind: both entangle mind and physical process already at the
ontological level. But the idea of the classical identity theory of the mind
is to hang onto the classical conception of physical reality, and aver that
a correct understanding of the true nature of a conscious thought would reveal
it to be none other than a classically describable physical process that
brings about what the thought intends, given the appropriate alignment of the
relevant physical mechanisms. 

That idea is, in fact, what would naturally emerge from  quantum theory
in the classical limit where the difference between Planck's constant and zero
can be ignored, and the positions of particles and their conjugate momentum 
can both be regarded as well defined, relative to any question that is posed. 
In that limit there is no effective quantum dispersion caused by the Heisenberg
uncertainty principle, and hence no indeterminism, and the only Heisenberg
Choices of questions about a future state that can get an answer `Yes' are
those that are in accord with the functional properties of the present state.
So there would be, in that classical approximation to the quantum process
described above, a collapse of the two lines of causation, the physical and
the mental, into a single one that is fixed by the local classical
deterministic rules. Thus in the classical approximation the mental process
would indeed be doing nothing beyond what the classical physical process is
already doing, and the two process might seem to be the same process. But
Planck's constant is not zero, and the difference from zero introduces
quantum effects that separate the two lines of causation, and allow their
different causal roles to be distinguished.

The identity theory of mind raises puzzles. Why, in a world composed
primarily of ontologically independent micro-realities, each able to access
or know only things in its immediate microscopic environment, and each
completely determined by micro-causal connections from its past, should
there be ontological realities such as conscious thoughts that can grasp
or know, as wholes, aspects of huge macroscopic collections of these
micro-realities, and that can have intentions pertaining to the future
development of these macroscopic aspects, when that future development is
already completely fixed, micro-locally, by micro-realities in the past?

The quantum treatment discloses that these puzzles arise from the conflation
in the classical limit of two very different but interlocked causal processes,
one micro-causal, bound by the past, and blind to the future,
the other macro-causal, probing the present, and projecting to the future.

\vspace{.2in}
\begin{center}
{\bf              Mental Force and the Volitional Brain        }
\end{center}
\vspace{.1in}

The psychiatrist Jeffrey Schwartz (1999) has 
described a clinically successful technique for treating patients with
obsessive compulsive disorder (OCD). The treatment is based on a program
that trains the patient to believe that his own {\it willful redirection} 
of his attention away from intense urges of a kind associated with 
pathological activity within circuitry of the basal ganglia, and toward 
adaptive functional behaviours, can, with sufficient persistent effort, 
systematically change both the intrusive, maladaptive, obsessive-compulsive 
symptoms, as well as the pathological brain activity associated with them.
This treatment is in line with the quantum mechanical understanding of 
mind/brain dynamics developed above, in which the mental/experiential 
component of the causal structure enters brain dynamics via intentions that 
govern attentions that influence brain activity.

According to classical physical theory ``a brain was always going 
to do what it was caused to do by local mechanical disturbances,''  
and the idea that one's ``will'',  is actually able to cause anything at 
all is ``a benign user illusion''. Thus Schwartz's treatment amounts, 
according to this classical conceptualization, to deluding the patient 
into believing a lie: according to that classical view Schwartz's 
intense therapy causes directly, in the patients behaviour, a mechanical 
shift that the patient delusionally believes is the result of his 
own {\it intense effort} to redirect his activities, for the purpose 
of effecting an eventual cure, but which (felt effort) is actually only 
a mysterious illusionary by-product of his altered behaviour.

The presumption about the mind/brain that is the basis of Schwartz's 
successful clinical treatment, and the training  of his patients, is that 
willful redirection of attention is efficacious. His success does not prove 
that `will' is efficacious, but it does constitute prima facie evidence that
it is. In fact, the belief that our thoughts can influence our 
actions is so basic to our entire idea of ourselves and our place 
in nature, and is so essential to our actual functioning in this world, that 
any suggestion that this idea is false would become plausible only under 
extremely coercive conditions, such as its incompatibility with basic physics. 
But no such coercion exists. Contemporary physical theory does allow our 
experiences, per se, to be truly efficacious and non-reducible: our 
experiences are elements of the causal structure that do necessary things 
that nothing else in the theory can do. Thus science, if pursued with 
sufficient care, demands no cloistering of disciplines, or interpretation 
as user illusions of the apparent causal effects of our conscious 
thoughts upon our physical actions.

\newpage
\begin{center}
 {\bf References}
\end{center}

Niels Bohr (1934), {\it Atomic Theory and the Description of Nature},
Cambridge Univ. Press, Cambridge.

Niels Bohr (1936), Causality and Complementarity, {\it Philos. of Science},
{\bf 4}. (Address to Second International Congress for the Unity of Science,
June, 1936).

Niels Bohr (1951), in reference A. Einstein (1951).

Niels Bohr (1958), {\it Atomic Physics and Human Knowledge}, Wiley,
New York.

C.B. Chiu, E.C.G. Sudarshan, and B. Misra, (1977) Phys. Rev. D {\bf 16},
520.

Daniel Dennett (1994), in {\it A Companion to the Philosophy of Mind},
ed. Samuel Guttenplan, Blackwell, Oxford. ISBN 0-631-17953-4.

A. Einstein (1951), {\it Albert Einstein: Philosopher-Physicist}
ed, P.A. Schilpp, Tudor, New York.

W. Itano, D. Heinzen, J. Bollinger, D. Wineland, (1990), Phys. Rev. {\bf 41A},
2295-2300.

Heisenberg, W. (1958a) `The representation of nature in contemporary 
    physics',  {\it Deadalus} {bf 87}, 95-108.

Heisenberg, W. (1958b) {\it Physics and Philosophy} (New York: Harper and 
     Row).

William James (1910), Some Problems in Philosophy, Ch X; in
{\it William James/ Writings 1902-1910}, The Library of America,
New York. (1987). ISBN 0-940450-18-0.

William James (1890), {\it The principles of Psychology, Vol I},
Dover, New York. ISBN 0-486-20381-6.

E. Joos (1986), {\it Annals, NY Acad. Sci. Vol 480 6-13.}
ISBN 0-89766-355-1.

Richard Rorty (1979), {Philosophy and the Mirror of Nature.}
Princeton U.P.

Jeffery M. Schwartz, A Role for Volition and Attention in the Generation
of New Brain Circuitry: Toward a Neurobiology of Mental Force, 
Journal of Consciousness Studies, June/July 1999.

Henry P. Stapp (1972), The Copenhagen Interpretation, Amer. J. Phys.
{\bf 40}, 2098-1116. Reprinted in Stapp (1993).

Henry P. Stapp (1993), {\it Mind, Matter, and Quantum Mechanics},
Springer-Verlag, New York, Berlin, Heidelberg. ISBN 0-387-56289-3. 

Henry P. Stapp (1998a), Pragmatic approach to consciousness, in
{\it Brain and Values: Is a Biological Science of Values Possible?}
ed. Karl H. Pribram, Lawrence Erlbaum, Mahwah, NJ. ISBN 0-8058-3154-1.\\ 
Or at www-physics.lbl.gov/(tilde)stapp/stappfiles.html
[(tilde) means the tilde sign] New Book ``Knowings'' (Book1.txt).

Henry P. Stapp (1998b), at www-physics.lbl.gov/\~{}stapp/stappfiles.html\\
See ``Basics'' for mathematical details about the vN/W formalism.

Steven Pinker (1997), {\it How the Mind Works}, Norton, NY.
ISBN 0-39304545-8.

W. Tittel, J. Brendel, H. Zbinden, and N. Gisin (1998), Physical Review
Letters {\bf 81}, 3563-3566.

J. von Neumann (1932), {\it The mathematical principles of quantum mechanics},
Princeton U.P. Princeton NJ, 1955.

Wigner, E. (1961) `The probability of the existence of a self-reproducing
      unit', in {\it The Logic of Personal Knowledge} ed. M. Polyani
     (London: Routledge \& Paul) pp. 231-238. 

W. Zurek (1986) {\it Annals, NY Acad. Sci. Vol 480, 89-97.}
ISBN 0-89766-355-1.

\end{document}